\documentclass[journal]{IEEEtran}
\ifCLASSINFOpdf
\else
\fi
\usepackage{soul}
\usepackage{graphicx}
\usepackage{subcaption}
\usepackage{algorithm,algorithmic}
\usepackage{amsmath}
\usepackage{multirow}
\usepackage{breqn}
\usepackage[table]{xcolor}
\definecolor{skyblue}{rgb}{0.53, 0.81, 0.92}
\definecolor{flax}{rgb}{0.93, 0.86, 0.51}
\usepackage{tabularx,booktabs}
\usepackage{amssymb}
\usepackage{algorithm,algorithmic}
\newcolumntype{C}{>{\centering\arraybackslash}X} 
\usepackage{lipsum}

\begin{document}
%
\title{Motion Classification using Kinematically Sifted  ACGAN-Synthesized  Radar Micro-Doppler Signatures}
%
%
%

\author{Baris Erol,~\IEEEmembership{Member,~IEEE,}
        Sevgi Zubyede Gurbuz,~\IEEEmembership{Senior Member,~IEEE,}
        Moeness G. Amin,~\IEEEmembership{Fellow,~IEEE}
\thanks{B. Erol is currently with Siemens Corporate Technology, Princeton, NJ.  Formerly, he was with the Center for Advanced Communications, Villanova University, PA, email: baris.erol@siemens.com.}
\thanks{S.Z. Gurbuz is with the Department
of Electrical and Computer Engineering, University of Alabama, Tuscaloosa,
AL, 30332 USA e-mail: szgurbuz@ua.edu.}
\thanks{M.G. Amin is with the Center for Advanced Communications, Department of Electrical and Computer Engineering, Villanova University, PA, email:  moeness.amin@villanova.edu.}
}

%
%

\markboth{Accepted to IEEE Transactions on Aerospace and Electronic Systems}{SKM: Aerospace and Electronic Systems} 

%



\maketitle

\begin{abstract}
Deep neural networks (DNNs) have recently received vast attention in applications requiring classification of radar returns, including radar-based human activity recognition for security, smart homes, assisted living, and biomedicine. However, acquiring a sufficiently large  training dataset remains a daunting task due to the high human costs and resources required for radar data collection. In this paper, an extended approach to adversarial learning is proposed for generation of synthetic radar micro-Doppler signatures that are well-adapted to different environments. The synthetic data is evaluated using visual interpretation, analysis of kinematic consistency, data diversity, dimensions of the latent space, and saliency maps. A principle-component analysis (PCA) based kinematic-sifting algorithm is introduced to ensure that synthetic signatures are consistent with physically possible human motions.  The synthetic dataset is used to train a 19-layer deep convolutional neural network (DCNN) to classify micro-Doppler signatures acquired from an environment different from that of the dataset supplied to the adversarial network.  An overall accuracy 93\% is achieved on a dataset that contains multiple aspect angles (0$^{\circ}$, 30$^{\circ}$, and 45$^{\circ}$ as well as 60$^{\circ}$), with 9\% improvement as a result of kinematic sifting. 

\end{abstract}

\IEEEpeerreviewmaketitle
\section{Introduction}
Over the past decade, radio frequency (RF) sensing has gained increased attention as its efficacy and unique advantages have been demonstrated for a variety of automotive, smart home, human computer interaction, and remote health monitoring applications 
 \cite{ahmad_signal_2016}-\nocite{pallotta_pseudo-zernike_2014,molchanov_ground_2011,mobasseri_time-frequency_2009,kim_human_2009,wu_radar-based_2015,su_doppler_2015}\cite{amin_radar_2017}.  Radar systems are both low-cost and low-power, making them a safe sensing alternative which can operate in darkness and all weather conditions. Moreover, radar is non-invasive, and when used for monitoring, does not require an alteration in daily habits or routines. These attributes have made RF sensing popular in motion monitoring.

Meanwhile, progress in machine learning and Internet of Things (IoT) is rapidly growing the expectations and performance requirements of ubiquitous sensing. Radar-based gesture recognition for man-machine interfaces requires an ability to recognize slight differences in hand motions, separating gestures intended to give commands versus daily hand movements \cite{soli}.  Biomedical applications of abnormal gait analysis, fall detection, fall risk assessment, and monitoring of hip/knee operations or neuro-muscular disorders, also require high sensitivity and specificity, consistent with medical standards \cite{Gurbuz2017,Seifert2017,Seifert2019,SeyfiogluCAE2018}.  Thus, even slight increases in accuracy and robustness are considered significant in the advancement of indoor radar technology and its adoption in smart homes and medical diagnosis.  

Deep neural networks (DNNs) have shown great potential to achieve high accuracy, even as the number of classes increases, and may well lead the way as a preferred method for motion classification in the near future \cite{8010417}-\nocite{trommel_multi-target_2016,kim_human_2016,chen_personnel_2018,kwon_human_2017}\cite{seyfioglu_deep_2017}. Yet, DNN architectures in RF applications are often limited by the fact that only small datasets are available for training, as data acquisition can be time consuming, costly, and limited in terms of the scope of scenarios and targets sampled.  This impacts not only DNN depth, but also the ability of the DNN to generalize across different body types, speeds, and motion classes \cite{Yang2019}, as well as adapt to different noise sources and environmental conditions.

Researchers have attempted to overcome this challenge by data augmentation, where the available radar data is modified through operations such as translation, time-shifting, and segmentation \cite{Ding2016,Wang2019}.  However, in RF applications, these approaches may not necessarily lead to statistically independent training samples that effectively span probable variations in target signatures. This is because the pixel values in the two dimensional (2D) data representations, generated through time-frequency (TF), analysis are related to the complex electromagnetic scattering and kinematics of the dynamic target being observed.  Radar returns from a moving target include not only a central Doppler shift, resulting from translational motion, but also micro-Doppler frequencies induced by slight rotations or vaibrations of parts of the target \cite{chen2011micro,chen2014radar}. In humans, micro-Doppler frequencies derive  from the unique, bipedal, time-varying kinematics of human motion, and varies even for the same activity depending upon body size, speed and individual gait style. Thus, methods for data augmentation motivated by image processing applications, such as scaling and rotating, may significantly disrupt RF data patterns by generating samples that are kinematically untenable.  The inclusion of such physically impossible samples in the training data has adverse effects, and compromises rather than improves performance.

To overcome these limitations, a simulation methodology rooted in kinematic modeling \cite{Ram2008,Ram2010,ErolMAES2015} via motion capture was recently proposed in \cite{seyfioglu_diversified_2018}. Instead of applying pixel-based data augmentation, transformations to the underlying skeletal model were applied to generate a large number of unique but kinematically consistent micro-Doppler signatures spanning expected target profiles.  A key disadvantage, however, is that the approach does not provide a means to account for the variations in signal-to-noise ratio (SNR), artifacts of sensor-related electronic interference, signal dispersion caused by frequency dependent obstructions, like walls, or non-target related motion (e.g., spinning ceiling fan).

Generative adversarial networks (GANs) have been proposed for synthesizing realistic images in a variety of applications \cite{GANoverview}, including synthetic aperture radar (SAR)  \cite{Lewis2018}.  An early effort at applying adversarial learning to synthetic data generation of micro-Doppler  was first proposed in 2018 \cite{Shi2018}, in which a Deep Convolutional GAN (DCGAN) was used to generate synthetic data that emulated the Boulic walking model.  The Boulic model consists of mathematically well-defined trajectories and therefore does not represent the spectral richness and intricacy of actual, measured micro-Doppler signatures.  By using such pristine and systematic simulated data to drive the DCGAN, replicas of the data were easily generated and nearly identical.

In another study \cite{Mi2018}, 150 simulated spectrograms were augmented with 1,000 GAN-generated spectrograms to classify a test set of 50 simulated signatures comprised of three activity classes: running, walking and jumping.  A 4\% increase in classification accuracy was noted. However, only a small number of samples were generated by the GAN, and the classes considered are easily identifiable so that the simulation study was not designed to vet the validity, merits or detriments of using GANs to simulate micro-Doppler signatures.

The first study exploiting adversarial learning for the classification of real micro-Doppler data was published in  \cite{Yang2019}, where Yang, et. al evaluated the efficacy of adversarial learning for addressing the open-set problem - the case where the training dataset does not include all the classes as the test dataset. Subsequent studies in \cite{ErolGAN2019_RadarCon} and \cite{KimGAN2019} utilized GANs for mitigating the problem of low sample support and reported the classification accuracy of DNNs trained with GAN-generated synthetic data for human activity recognition.

In fact, the ability of GANs to synthesize authentic radar micro-Doppler signatures is hampered by differences between radar phenomenology and optics.  The values of pixels in micro-Doppler signatures relate not to physical shapes, but instead to human kinematics.  It is thus possible for GANs to generate numerous synthetic samples that, while visually similar, are incompatible with the kinematics of human motion.

The work in this paper is developed concurrently with that of \cite{ErolGAN2019_RadarCon}\cite{KimGAN2019} and provides, to our knowledge, the first in-depth analysis of GAN-generated synthetic data in terms of kinematic fidelity and diversity. In particular, we propose the utilization of auxillary classifier generative adversarial networks (ACGANs) \cite{odena_conditional_2016}, as opposed to
conditional variational autoencoders (CVAEs) \cite{doersch_tutorial_2016}, for the generation of synthetic micro-Doppler signatures with greater diversity and sharpness. The issues of kinematic fidelity of the ACGAN-generated synthetic data are illustrated using physics-based rules applied to walking and falling motion classes.  The relationship between kinematic fidelity and the dimensionality of the latent space as well as sample diversity is also examined.  We propose a new technique for kinematic sifting based on principal component analysis (PCA) to eliminate the kinematically impossible samples from the synthetic training dataset and, as such, limit their corrupting effects on performance.  This underlies the importance of considering kinematics when generating synthetic micro-Doppler signatures using adversarial learning. The proposed technique achieves a 9\% improvement in performance over that attained if the ACGAN-generated signatures are used directly for training, without kinematic sifting.

Finally, we show the benefits of ACGAN-generated synthetic data to adaptation to different sensing locations and environments. A small number of measured radar data collected from one location (with multiple aspect angles: 0$^{\circ}$, 30$^{\circ}$, 45$^{\circ}$ and 60$^{\circ}$) is used by ACGAN to grow the synthetic dataset for training, while the test dataset is collected at a different location in a through-the-wall configuration. A 19-layer convolutional neural network (CNN) trained using the kinematically sifted data generated by the ACGAN is shown to yield a high 93\% classification accuracy across different environments.

The paper is organized as follows: In Section II, the experimental radar measurements conducted in two distinct locations and environments is described. In Section III, the generative model, ACGAN, is discussed in relation to an alternative generative model, CVAE. In Section IV, diversity, accuracy and kinematic fidelity of the ACGAN-generated synthetic images are evaluated. In Section V, classification with PCA-based kinematic sifting of ACGAN-generated synthetic data for training a 19-layer CNN in a scenario involving adaptation across two distinct environments is presented.  Discussion of conclusions and future work is provided in Section VI.

\begin{figure*}[t!]
\centering
\includegraphics [width=6in,keepaspectratio] {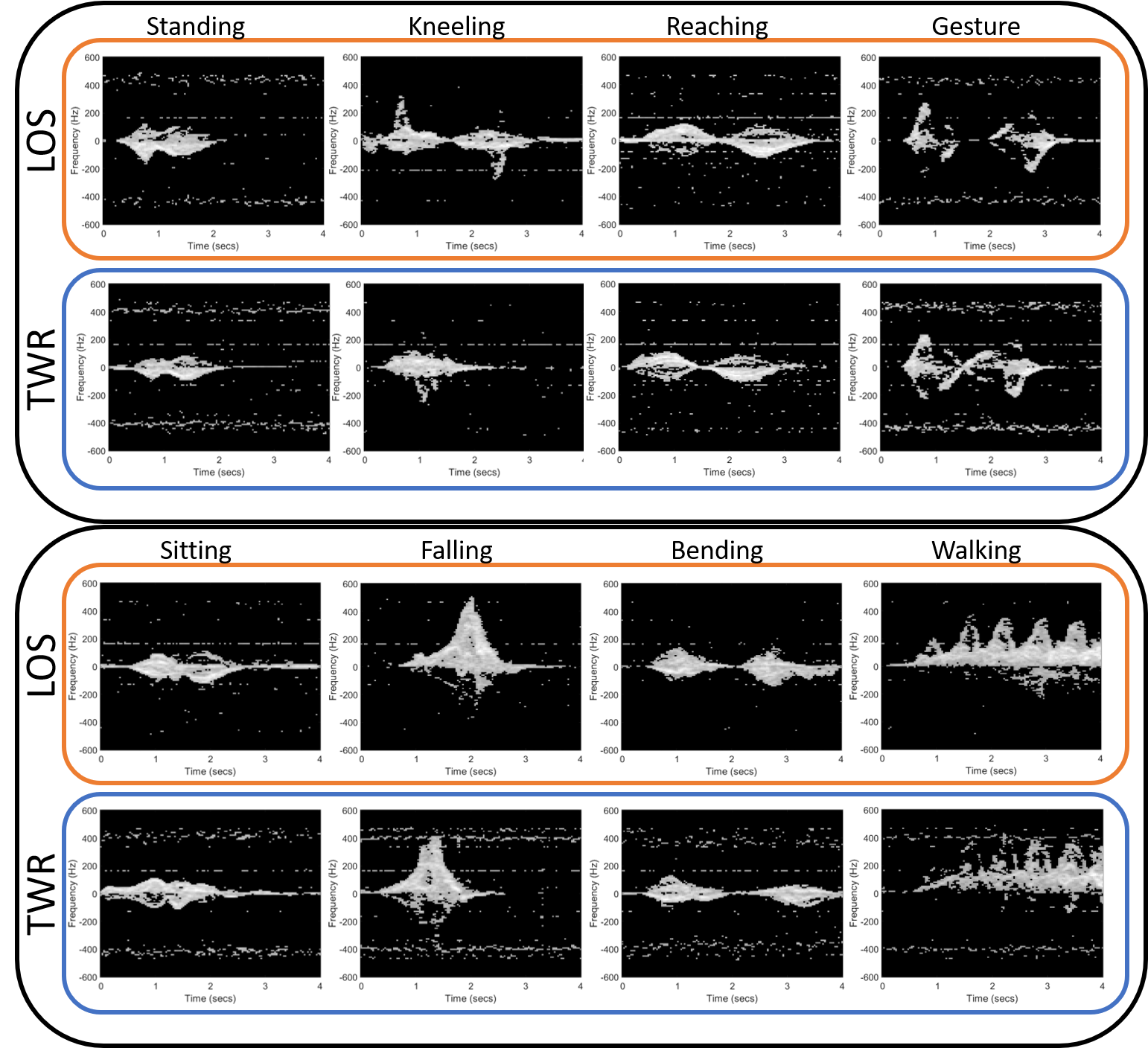}
\caption[ ]
{\small Real spectrogram images (after all pre-processing) of different human activities. } 
\label{fig:real_images}
\end{figure*}

	

\section{Radar Micro-Doppler Measurements}
Commercially available continuous-wave (CW) radars are compact in size and provide a measurement of Doppler frequencies as a function of time. The radar system used in this work operates at a transmitting frequency of 25 GHz, sweep time of 10 ms, while collecting 128 samples per sweep. Thus, the received radar signal is highly oversampled at a rate of 12.8 kHz \cite{noauthor_sdr-kit_nodate}. A higher sampling frequency causes the spectrum to shrink and cluster around the origin, leaving considerable vacant space in the time-frequency domain. Therefore, during pre-processing, the received radar signal is first downsampled to 1.2kHz. The power output and antenna gain of the radar are 16 dBm and 18 dBi, respectively.

\subsection{Time-Frequency Representation: Spectrograms}
Human activity recognition is typically accomplished through identification of unique patterns in the radar micro-Doppler signature, a time-frequency representation of the radar received signal \cite{boashash_time-frequency_2015}. Quadratic time-frequency (QTF) distributions are considered a powerful tool for the analysis of time-varying signals, with spectrograms being the simplest and most commonly used TF distribution \cite{sejdic_timefrequency_2009}. Spectrograms are the energetic form of the short-time Fourier transform (STFT), which is obtained by splitting the time domain signal into many overlapping or disjoint consecutive segments, and then taking the Fourier transform (FT) of each segment. A spectrogram thus exposes the signal's local frequency behavior and is mathematically defined as
\begin{equation}
\mathrm{S}(n,k)=\bigg |\sum_{m=0}^{N-1} \mathrm{h}(m)\mathrm{x}(n-m)e^{-j2\pi km/N}\bigg|^2,
\end{equation}
where $\mathrm{h}(m)$ is a window function which can affect both the time and frequency resolutions. The window slides over the data to capture the instantaneous frequencies.  The amount of overlap is variable, so that the window could slides one or more samples each time. At each window time-position, the local frequency behavior is emphasized through the windowed Fourier transform (FT). The window length trades off spectral and temporal resolutions, with long windows providing high frequency resolution, whereas short windows offer high temporal resolution.

Optimal sampling frequency and STFT parameters can be found using a grid search; however, this might not lead to a global optimum since the parameter step size is determined manually. Therefore, we used data-driven optimization with genetic algorithms (GA) to determine the optimum hyperparameters of the STFT and the sampling frequency, while maximizing the classification performance achieved by generalized PCA (GPCA) and minimum distance classifier (MDC). For one set of hyperparameters, GPCA is used to reduce the dimensionality and extract features in time and frequency, which are subsequently provided to the MDC. Classification accuracy is used as the fitness function of the GA, while the GA structure was selected as  NSGA-II -- one of the most popular multi-objective optimization algorithms \cite{deb_fast_2002}. 

The upper and lower bound of the hyperparameters are determined as: Sampling frequency 200Hz-12kHz, window length 64-1024 (in samples), overlapping length 64-1024 (in samples), number of FFT points 128-4096 (in samples). Only one constraint is forced into the optimization procedure: namely, that the window length must be greater than the overlapping length. 

Based on this approach, in this work, spectrograms are generated using 1024 frequency samples, a Hanning window of length 512, and an overlap of 256 samples. Note that after the spectrograms are computed, they are converted to grayscale prior to input to the ACGAN and CVAE. Before the pre-processing for the clutter mitigation, spectrograms were converted to grayscale and resized into $100 \times 100 \times 1$. This is the final dimensionality of the inputs provided to generative models and the 19-layer CNN.

\subsection{Experimental Datasets Collected}
In this work, eight (8) different activities are considered:

\begin{itemize}
\item Bending - person stands and moves torso from a vertical to horizontal position, resulting in both positive and negative frequency components over the same time interval.  Positive frequencies result from the forward movement of torso, coupled with negative frequencies due to the posterior moving away from the radar.
\item Falling - person falls forward onto a mattress, resulting in the shape of an upside-down bow in the signature.We only consider non-progressive falls which exhibit relatively high Doppler frequencies.
\item Gesturing - gross arm motion, such as by moving the arm up and down to turn the TV on/off, or pointing a lamp with different orientations to turn it on/off.
\item Standing - in-place motion of a person to standings up from the sitting position.
\item Kneeling - person lowers position to set one knee on the ground, as one would when tying shoelaces.  This results in a distinct spike in the micro-Doppler signature.
\item Reaching - person extends torso and arms upward from a sitting position.
\item Sitting - person is standing upright, then sits on a chair.
\item Walking - micro-Doppler signature exhibits a distinct sinusoidal pattern for the strongest return caused by the slight up-and-down motion of the torso incurred as a function of time. The periodic forward-backward motion of the arms and legs results in higher amplitude, periodic oscillations modulated around the main Doppler shift. Leg motion causes the highest frequency oscillation, followed by that of the arms, which appear at distinct, mid-level frequencies.
\end{itemize}

A sample spectrogram for each class collected in two different settings is shown in Figure  \ref{fig:real_images}.  To create an environment for radar measurements different for training and testing data, we placed the radar in an adjacent room with obstructed line-of-sight to the target through an interior wall.  The  line-of-sight (LOS) dataset was acquired from the Radar Imaging Laboratory, while the dataset associated with an obstructed LOS was acquired at the Center for Advanced Communications (CAC) conference room, both located at Villanova University. The latter sensing environment is meant to generate through-the-wall radar (TWR) dataset.  The radar was placed on a table with a height of 3.2 ft for both of the locations.  In the LOS experiment, a total 14 participants were involved in the data collection (12 male and 2 female), who had heights ranging from 5.1 to 6.3 ft, and weights ranging from 119 to 220 lbs. All activities were conducted for three different walking angles (0$^{\circ}$, 30$^{\circ}$, and 45$^{\circ}$) and three different speeds (slow, typical and fast), resulting in a dataset that covers a wide variety of motions with sufficient intra- and inter-class variance. A total of 1586 samples were collected, with the number of samples per class shown in parenthesis as follows: bending (167), falling (350), kneeling (216), gesture (150), reaching (140), sitting (233), standing (130), walking (200). 

In contrast with the LOS dataset, the radar and test subject were separated by a plywood wall. Subjects started the motion 5 meters away from the wall and after 4 seconds of data collection experiment is repeated. Experiments included both moving towards and away from the radar. The TWR dataset was conducted at four different angles, including 0$^{\circ}$, 30$^{\circ}$, and 45$^{\circ}$ as well as 60$^{\circ}$.  The test subject in the TWR experiments was a male participant, who was not part of the LOS data collect.  A total of 387 TWR samples were collected, with bending (50), falling (72), gesture (50), kneeling (15), reaching (50), sitting (50), standing (50) and walking (50). A summary of the LOS and TWR datasets is given in Table \ref{tab:dataset}.

In this work, the LOS dataset was used in conjunction with the ACGAN for training data generation, while the TWR dataset was used for testing. Note that the visual similarity between 7 of 8 classes (walking is the exceptional class), inclusion of multi-angle measurements, and difference in environment makes this classification problem relatively more challenging \cite{narayanan} in comparison to other scenarios considered in the literature.

\begin{table*}[t!]
	\begin{center}
	\caption{Experimental Dataset Summary}
	\label{tab:dataset}
	\begin{tabular}{|c|c|c|c|c|c|}
		\hline
		 & \textbf{$\#$ subjects} & \textbf{Aspect angles} & \textbf{$\#$ activities} & 
        \textbf{Location} & \textbf{$\#$ samples} \\
		\hline\hline
		LOS & 
        14 & 0$^{\circ}$, 30$^{\circ}$, 45$^{\circ}$ &
        8 & CAC Conf. Room       &
        1586 \\ \hline
		TWR & 1* & 0$^{\circ}$, 30$^{\circ}$, 45$^{\circ}$, 60$^{\circ}$ & 8 & Radar Imag. Lab & 387  \\
		\hline
	\end{tabular} \\

* Subject different from those in LOS dataset.
	\end{center}
	\label{tab:test7class}
\end{table*}

\subsection{Pre-processing for Clutter Mitigation}
The classical signal processing approach to deal with environmental factors is to remove any clutter or unwanted artifacts using filtering.  In this work, we applied an approach known as the extended CLEAN (eCLEAN) algorithm, which was originally designed for range-Doppler processing \cite{8461512}. eCLEAN aims at suppressing unwanted distortions or noise effects while enhancing the natural structural integrity of the data. Simple pre-defined thresholding is the most commonly pre-processing method in micro-Doppler processing. However, due to high variance in our data (different aspect angles, data collection environments, subjects etc.), determining a threshold that works for every data/class is challenging. A simple example is provided in Figure 2 for a walking micro-Doppler image filtered with (a) simple thresholding and (b) eCLEAN algorithm. Note that, this threshold value works really well for some of the other walking micro-Doppler images, however, for this particular walking example it did not remove any of the noise components, which would degrade classification performance. On the other hand, eCLEAN removes all the noise and artifacts without needing a pre-defined threshold. It automatically determines the number of points which are needed to be removed using a simple and efficient histogram-based method. eCLEAN firstly computes the 2D histogram of the sample spectrogram, dowmsamples it and applies a normalization. Afterwards, it automatically determines the threshold where the number of counts is below 0.1. After the threshold is acquired, it slides over the time axis and examines each time column and determines the number of points should be extracted depending on the threshold. It operates on time slices and creates mask functions. This continues until all number of points are extracted. An example pseudocode of the eCLEAN is provided in Algorithm 1.

\begin{algorithm}[b!]
	\caption{eCLEAN algorithm}
	\begin{algorithmic}[1]
		\renewcommand{\algorithmicrequire}{\textbf{Input:}}
		\renewcommand{\algorithmicensure}{\textbf{Output:}}
		\REQUIRE Training spectrogram datacube ($\mathcal{X} \in \mathbb{R}^{I_1 \mathrm{x} I_2 \mathrm{x} I_3}$, $I_1$ and $I_2$ original image sizes and $I_3$ number of training samples),   \\
		\ENSURE  Cleaned training spectrogram datacube ($\mathcal{X}_c \in \mathbb{R}^{I_4 \mathrm{x} I_5 \mathrm{x} I_3}$, $I_4$ and $I_5$ resized spectrogram dimensions)\\ 
	\textbf{PROCESS:} \\
		\FOR {$n=1$ to $I_3$}
		\STATE $\mathrm{Y} \in \mathbb{R}^{I_1 \mathrm{x} I_2} \leftarrow \mathcal{X}(:,:,n), $ matrix slice of tensor  $\mathcal{X}$
		\STATE Normalize \& resize the matrix $\mathrm{Y}$ and compute 2D histogram 
		\STATE Find the intensity index ($\alpha_n$) where distribution starts to 
        plateau 
        \FOR {$k=1$ to $I_2$}
        \STATE $\mathrm{p} \in \mathbb{R}^{I_1} \leftarrow \mathrm{Y}(:,k) $, fiber of tensor  $\mathcal{X}$
        \STATE Compute 1D histogram of fiber $\mathrm{p}$
        \STATE Apply $\alpha_n$ and determine the number of points ($\mathrm{N}_s$) should be extracted from fiber $\mathrm{p}$
                \FOR {$j=1$ to $N_s$}
        	\STATE $f_j = \max_k (\mathrm{p} \in \mathbb{R}^{I_1}), \ \ v_j = \arg\max_k (\mathrm{p} \in \mathbb{R}^{I_1})$
            \STATE Subtract a fraction of the point spread function centered at the peak from $\mathrm{p}$. 
            \STATE Record the peak amplitude and position in the cleaned vector ($\mathrm{p}_c \in \mathbb{R}^{I_1}$)
		\ENDFOR
        \STATE Store $\mathrm{p}_c$'s in cleaned spectrogram image $\mathrm{Y}_c \in \mathbb{R}^{I_1 \mathrm{x} I_2}$
        \ENDFOR
		\STATE Store $\mathrm{Y}_c$'s in cleaned spectrogram datacube $\mathcal{X}_c$
		\ENDFOR
	\end{algorithmic}
	\label{tab:eclean}
\end{algorithm}

\begin{figure}[t!]
	
\centering
\includegraphics[width=3.5in, keepaspectratio]{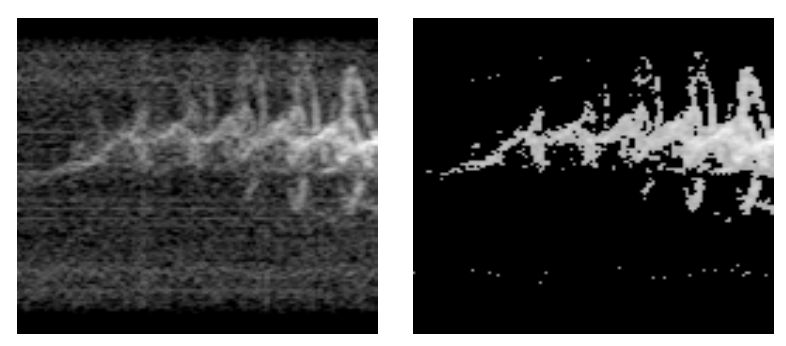}
	\caption{Preprocessing of micro-Doppler images (a) Pre-defined thresholding (b) Proposed eCLEAN.}
	\label{fig:Kinematic_Fig_1}
	
\end{figure}

All spectrograms illustrated in Figure \ref{fig:real_images} have had the eCLEAN algorithm applied on the data.  Thus, it is important to note that clutter mitigation was not sufficient in removing all artifacts in the data, and that environmental differences remain in the two datasets despite such mitigation efforts.  This point is significant because it underscores the necessity of developing DNN approaches that can overcome non-target artifacts present in the data.



\section{Generative Models}
The term "generative" is used in many ways in the machine learning community. Within the scope of this work, this term refers to a model that takes a training data with distribution $p_{data}$ and seeks to learn a close estimate of it, denoted as $p_{model}$. More specifically, generative models attempt to predict features given a certain label, whereas, discriminative models try to predict a label of a given input data \cite{salimans_improved_2016}-\nocite{theis_note_2015}\cite{radford_unsupervised_2015}. Generative models can be classified into two broad categories: explicit (VAE, PixelRNN/CNN \cite{oord_conditional_2016,oord_pixel_2016}) and implicit (GAN \cite{goodfellow_generative_2014}, Markov chain) approaches \cite{goodfellow_nips_2016}.

Generative models have been successfully employed in image recognition, such as performance improvement in reinforcement learning, domain adaptation, presentation and manipulation of high dimensional distributions and overcoming the problems with missing data \cite{goodfellow_nips_2016}. In this work, we apply generative models in the context of human motion classification to increase the amount of training data as well as to broaden the intra-class motion diversity, while taking into account environmental factors, e.g., clutter sources, which are not included in kinematic models of human motion.  The Wasserstein GAN (WGAN) is a popular variant of the GAN architecture, which employs the 1-Wasserstein distance, also known as the Earth-Mover (EM) distance rather than alternative metrics, such as the Kullback-Leibler (KL) divergence or the Jenson-Shannon Divergence (JSD), to quantify the distance between the model and target distributions \cite{wgan}.  The WGAN is advantageous because it provides for a more stable training process, with proven convergence of the loss function, and is less sensitive to model architecture or hyperparameter selection.

\begin{figure*}[t!]
\centering
\includegraphics [width=5in] {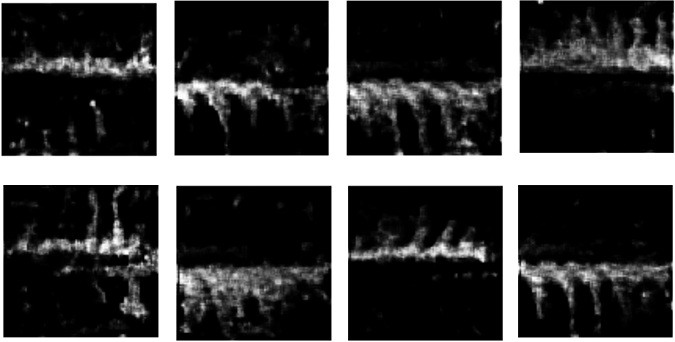}
\caption{Randomly chosen 8 samples generated by  WGAN for walking class.}
\label{fig:wgan}
\end{figure*}

The results of applying a WGAN to synthesize radar micro-Doppler signatures for walking is shown in Figure \ref{fig:wgan}.  It may be observed that many of these samples have features that are deviant from the typical properties of walking micro-Doppler, such as high frequency components disconnected from the low-frequency micro-Doppler of the torso, negative micro-Doppler corresponding to motion in the reverse direction, and filled in regions between the peaks that would be inconsistent with the arm motion of a typical walking person.

As a result, in this work, we focus on conditional generative models, principally the CVAE and ACGAN, which allow the generative model to condition on external class labels. This has the benefit of improving the visual accuracy of the synthetic images generated. An alternative to these conditional models is to train $N$ (number of classes) separate models. However, this has an adverse effect of causing the problem of overfitting due to the small amount of available training data, and requires high computational power. Moreover, it has been shown that forcing a model to perform additional tasks or constraints improves the performance of the original problem \cite{odena_conditional_2016}.

\subsection{Conditional Variational Autoencoders (CVAE)}
CVAEs are an extension of the vanilla VAE where the input observations modulate the prior on Gaussian latent variables that generate the outputs \cite{sohn_learning_2015}. A vanilla VAE consists of an encoder, a decoder, and a loss function. The encoder and decoder are usually designed as neural networks, and they are given the weights of $\theta$ and $\phi$, respectively. The encoder takes an input image and outputs a latent representation in lower dimensions.  It is important to note that the latent space is stochastic: the encoder outputs parameters to a Gaussian probability density, which can then be sampled to obtain noisy values of the latent representation $\mathrm{z}$. Then, the decoder takes the encoded latent representation as an input and outputs parameters to the probability distribution of the data.  In this work, the encoder and decoder are denoted as $q_{\theta}(\mathrm{z}|\mathrm{x})$ and $p_{\phi}(\mathrm{x}|\mathrm{z})$, respectively.

The loss function of a vanilla VAE is the negative log-likelihood with a regularizer. It can be decomposed into a single spectrogram image since there are no global connections between images. The loss function $l_i$ for a single image $\mathrm{x}_i$ is defined as
\begin{equation}
l_i(\theta, \phi) = - \mathrm{E}_{\mathrm{z} \thicksim q_{\theta}(\mathrm{z}|\mathrm{x}_i)}[\log p_{\phi}(\mathrm{x}_i|\mathrm{z})] + \mathrm{KL}(q_{\theta}(\mathrm{z}|\mathrm{x}_i)||p(\mathrm{z}))
\end{equation}
where the first and second term represent the reconstruction error and the regularizer, respectively. The former encourages the decoder network to learn how to reconstruct the input data, while providing the smallest error, as in basic autoencoders. If the decoder is unable to reconstruct the data well enough, then it will incur a high loss function value. The regularizer is the Kullback-Leibler (KL) divergence, which measures how much information is lost when using $q_{\theta}(\mathrm{z}|\mathrm{x})$ to represent $p(\mathrm{z})$. The regularization term forces the encoder to map images from the same classes onto the same region in the latent space. Moreover, in the VAE, $p$ is specified as the normal distribution with mean zero and variance one ($N(0,1)$).

Similar to vanilla VAEs, a CVAE consists of an encoder, a decoder, and a loss function. However, in contrast to VAEs, CVAEs have additional input branches called conditions (external class labels) to both the encoder and decoder. Due to embedding of class labels, the encoder is conditioned on the spectrograms and corresponding class labels, whereas the decoder is conditioned on latent variables and class labels. Other than conditional embeddings, CVAEs have the same principle as VAEs, where the encoder takes the spectrograms and class labels ($\mathrm{x}, \mathrm{y}$) and outputs a hidden representation $z$, with the attached weights ($\theta$) and biases ($\phi$).  Then, the decoder takes $z$ and $\mathrm{y}$ as inputs and outputs the parameters to the probability distribution of the data. The CVAE is trained to maximize the conditional log-likelihood. In CVAEs, the empirical lower bound is defined as
\begin{multline}
L_{cvae}(\mathrm{x},\mathrm{y};\theta,\phi) = -\mathrm{KL}(\ q_{\phi}(\mathrm{z}|\mathrm{x},\mathrm{y}) \ || \ p_{\theta}(z|\mathrm{x})) \\
+ \frac{1}{L} \sum_{l=1}^{L} \log p_{\theta} (\mathrm{y}|\mathrm{x},\mathrm{z}^{(l)}),
\end{multline}
where $\textbf{z}^{(l)}\approx N(0,1)$, $L$ is the number of samples, $q_{\phi}(z|x,y)$ is the conditional recognition distribution, and $p_{\theta}(z|x)$ is the generative distribution. A more detailed theoretical background and implementation considerations on VAE and CVAE can be found in \cite{doersch_tutorial_2016}.


As a pre-processing step, the input spectrograms ($64 \times 64 \times 1$) are reshaped into flat vector representations of $4096 \times 1$ pixel values. Then, the vectorized spectrogram images and class labels are concatenated. In our case, the input size of the CVAE is $4104 \times 1$ (reshaped image size + number of classes). The encoder and decoder configurations used in this work consist of fully-connected (dense) layers. The encoder takes the merged data and passes it to sequential dense layers with specified neurons and activation functions: ($2048 \times 1$) - ReLU, ($1024 \times 1$) - ReLU, ($512 \times 1$) - ReLU, and $10 \times 1$ - Linear. The encoder has a total of 11,018,762 (trainable) parameters. The final layer is responsible for the mean and standard deviation for the variational sampling that will occur from the latent space $\mathrm{z}$. After sampling, the decoder reconstructs $\hat{\mathrm{x}}$ and consists of four dense layers as ($512 \times 1$) - ReLU, ($1024 \times 1$) - ReLU, ($2048 \times 1$) - ReLU, and ($4096 \times 1$) - Sigmoid. The decoder has total of 11,022,848 (trainable) parameters. We applied stochastic gradient descent with Adam optimizer \cite{ADAM}, an adaptive moment estimation method, controlled by parameters $\beta_1 = 0.5$ and $\beta_2 = 0.999$. The learning rate is determined as 0.0005 for 500 epochs and minibatch size of 16. A total of 40,000 synthetic spectrograms are generated using CVAE (5000 for each class).

\subsection{Auxiliary Classifier Generative Adversarial Networks}

GANs are implicit generative models that aim to learn the data distribution from a set of training samples. Due to their implicit structure, generative models do not need any intractable density functions as in CVAE. The basic idea of GANs stems from a game-theoretic approach between two players (both neural networks): generator and discriminator. These two entities are in constant battle during training.  The generator ($\mathrm{G}$), seeks to generate samples that are intended to come from the same distribution of the training data. The input of the generator can be sampled from a Gaussian distribution as random noise. The generator gets samples $\mathrm{z}$ from the selected distribution and maps $G(z)$ to the image space. The main goal of the generator is to make the image space distribution as close as possible to the $p_{data}$. The second network is called discriminator and denoted as $\mathrm{D}$. The role of the discriminator is to discriminate between real and fake samples generated by the generator. It takes a simple input $\mathrm{x}$ and outputs $\mathrm{D}(\mathrm{x})$ which is a probability of the given image is being real.

Since GANs use a game-theoretic application, the objective function can be represented as a minimax function. In essence, the discriminator tries to maximize the objective function using gradient ascent, whereas the generator tries to minimize the objective function using gradient descent. Training of these networks can be done by alternating between gradient ascent and descent. The loss function of the adversarial networks can be shown as
\begin{equation}
\min_{G} \max_{D} \mathrm{E}_{x \thicksim p_{data}} \log(D(x) + \mathrm{E}_{z \thicksim p_z}[\log(1-D(G(z)))]).
\end{equation}

In the objective function, the discriminator is trained to maximize the $\mathrm{D}(\mathrm{x})$ for images with $\mathrm{x} \thicksim p_{data}$. The objective of the generator is to produce images $G(z)$ to fool $\mathrm{D}$ during training such that $\mathrm{D}(\mathrm{G}(\mathrm{z})) \thicksim p_{data}$. During training, the generator improves its ability to synthesize more realistic images while discriminator improves its ability to distinguish between real from fake images. 

ACGAN is an extension of the vanilla GAN model that enables the model to be conditioned on external labels to improve the quality of the generated images. One method to produce class conditional samples can be done by supplying both generator and discriminator with class labels as in CVAE. However, the strategy behind the ACGAN is to instead of feeding the class information to the discriminator, one can task the discriminator with reconstructing the label information. This can be done by modifying the discriminator to contain an auxiliary decoder network that outputs the class labels for the training data \cite{odena_conditional_2016}. In this respect, the objective function of the ACGAN has two parts: the log-likelihood of the correct source, $L_{\mathrm{s}}$, and the log-likelihood of the correct class, $L_{\mathrm{y}}$.
\begin{multline}
L_s = \mathrm{E}[\log p(\mathrm{s} = real | \mathrm{x}_{real})]\\
+ \mathrm{E}[\log p(\mathrm{s} = fake | \mathrm{x}_{fake})].
\end{multline}
\vspace{-0.3in}
\begin{multline}
L_{\mathrm{y}} = \mathrm{E}[\log p(\mathrm{Y} = y | \mathrm{x}_{real})]\\
+ \mathrm{E}[\log p(\mathrm{Y} = y | \mathrm{x}_{fake})],
\end{multline}
where $\mathrm{s}$ are the generated images. The discriminator is trained in order to maximize the $L_s + L_Y$ while the generator is trained to maximize $L_Y-L_s$.

The employed ACGAN architecture consists of two different parts: generator and discriminator. The generator takes a vector of $100 \times 1$ random noise (latent space) drawn from a uniform distribution ($N(0,2)$) and class labels as inputs and outputs a spectrogram image of size $64 \times 64 \times 1$. We used a similar generator network, as in the original ACGAN paper, with minor modifications for generating radar spectrogram images. The generator consists of a fully connected dense layer reshaped to size $4 \times 4 \times 128$ and four 2D convolutional layers with $3 \times 3$ kernel size. Filter sizes for each convolutional layer are determined as 256, 128, 64 and 1. The last layer contains only one filter due to gray-scale channel size. Batch normalization with the momentum of 0.8 and 2D up-sampling (kernel size $2\times 2$ with strides of 2) are  applied to each layer (including the dense layer) of the generator network, except for the output layer. In addition to the batch normalization, dropout of 0.15 is also applied in every even layer considering the small amount of real training data. Adding these regularizes into the generator network helps combat overfitting and mode collapsing. ReLU activation functions are applied to all convolutional layers except the output layer which employs a tanh activation function. Discriminator structure consists of seven 2D convolutional layers with a kernel size of $3\times3$. LeakyReLU is utilized as an activation function after every convolutional layer except for the last one (the slope of the leak was set to 0.2).  Max-pooling is only included in the first layer with a filter size of $2\times2$ and strides of 2. Downsampling is done in every odd convolutional layer with a stride rate of 2. Batch normalization with momentum 0.8 is utilized in every layer except for the first one.  In addition to batch normalization, a dropout of 0.25 is applied in every even layer. The number of filters in each convolutional layer is determined as 64, 128, 128, 256, 256, 512, and 512. The last layer of the discriminator uses a sigmoid for the validity of the generated images and softmax for reconstruction of the class labels.  

\begin{figure*}[t!]
\centering
\includegraphics [width=7.1in] {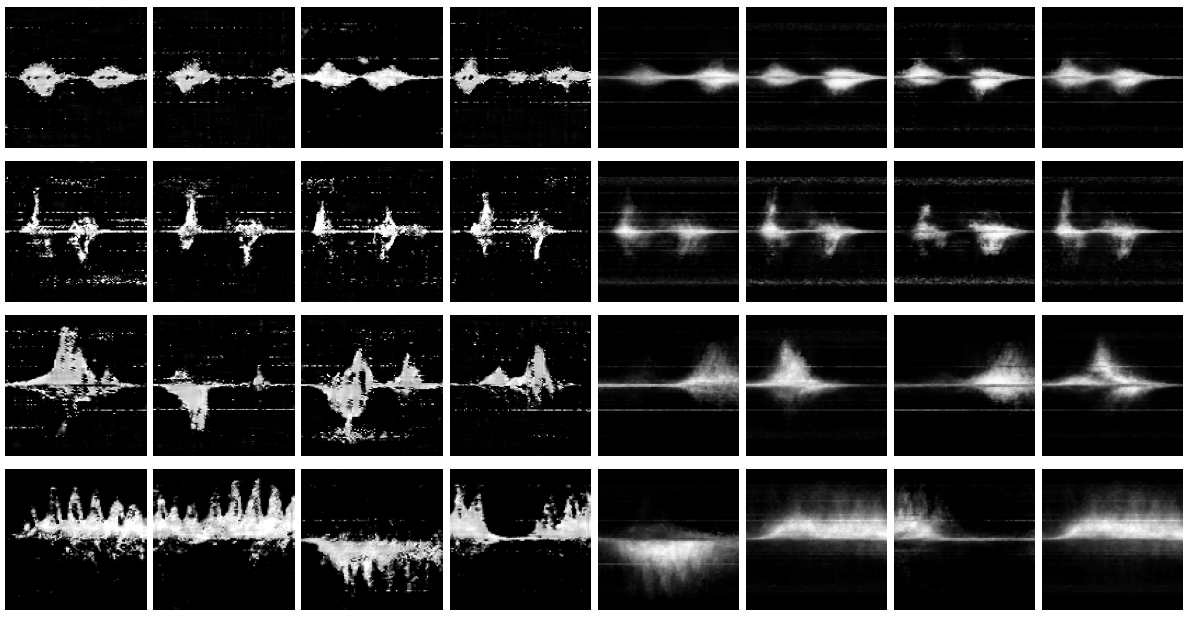}
\caption{Randomly chosen four samples generated by  ACGAN (first four in row-wise) and CVAE (last four in row-wise). Each row represents a different class (bending, gesture, falling and walking).}
\label{fig:generated_images}
\end{figure*}

The pre-processing step for the ACGAN (also for CVAE) includes a cleaning and filtering algorithm which is followed by scaling of the images between $(-1, 1)$ for tanh activation function. Weights are initialized with a normal distribution. An Adam optimizer is utilized with learning rate of 0.0002, $\beta_1 = 0.5$, and $\beta_2 = 0.999$ for 3000 epochs and minibatch size of 16. Some examples generated by the proposed ACGAN are depicted in Figure \ref{fig:generated_images}. A total of 40,000 synthetic spectrograms are generated using ACGAN (5000 for each class). 

Note that the training of CVAE and ACGAN are done offline with a PC equipped with GT 1080Ti. The computational cost of the CVAE is low due to the fast convergence (around 500 epochs) of the autoencoder topology. However, for the ACGAN, convergence takes more time due to the minmax structure of the adversarial learning. Moreover, the topologies of the generator and discriminator in the ACGAN are more complex than that of the CVAE, which results in increased computational time costs.  Our experimentation shows that the ACGAN converges after around 5000 epochs.

\section{Kinematic Evaluation of Synthetic Signatures}
Despite progress in the theoretical understanding of generative models and increased attention in GAN research, evaluating and comparing the performance of these models still remain a hard task. While several measures have been introduced, there is no consensus yet as to which measure best captures the strengths and limitations of the models and yield a fair model comparison \cite{borji_pros_2018}. Moreover, evaluation metrics are usually problem specific. Because the underlying physics of the problem is different, performance metrics valuable in the optical domain, such as inception score or Fr\'echet Inception Distance (FID), do not necessarily translate to the RF domain.

In radar micro-Doppler classification, important challenges in the context of training include obtaining a sufficient amount of real data to drive synthetic data generation with GANs, while ensuring that the synthetic signatures are diverse, spanning the characteristics of all expected motion signatures, and correspond to human motion that is physically possible.  In the underlying problem, the human skeleton constrains the possible variations of spectrograms corresponding to a given class - a condition we should observe to avoid erroneously training the network.  Towards this end, we consider four measures to evaluate the efficacy of the generative networks: 1) visual inspection, 2) kinematic fidelity, 3) signature diversity, and 4) the dimension of the latent space.

\subsection{Visual Inspection}
A sample of some of the spectrograms generated by CVAE and ACGAN are shown in Figure \ref{fig:generated_images} for four classes: bending, gesture, falling and walking. At the outset, it may be noticed that the CVAE-generated signatures are almost unrealistically blurry, a feature exhibited across all classes.  The main reason for this blurriness stems from the challenge of fitting of the data distribution into a tractable density distribution. 

\subsubsection{Bending} 
Both ACGAN and CVAE capture the most essential kinematic property of bending, namely, the presence of positive and negative frequency components within the same time limit. Moreover, ACGAN was able to learn to place time separation between the first and second part of the bending motion. In some generations, the time difference between the bending down (first hump) and standing up (second  hump) is close  0.2 seconds, whereas in some other generations it is up to 2 seconds. 

\subsubsection{Gesturing}
For gesturing, the ACGAN generated some variations capturing the different orientations and velocities of the arm. CVAE again seizes the kinematic property of the motion; however, generated images remain blurry. 

\subsubsection{Falling}
For falling, ACGAN underscored some salient features about the motion articulation. Note that, all real falling experiments in the LOS dataset were performed towards the radar, resulting in positive Doppler frequency. Interestingly, ACGAN learned how to mirror the spectrogram and generated some examples as if the subject had performed the motion in the opposite direction. Moreover, in some cases, ACGAN generated signatures that resemble "progressive falling": i.e., putting the knee first - holding onto something then falling. 

\subsubsection{Walking}
In walking, again the principal kinematic properties are captured by ACGAN. The motions of the legs and arms can be seen in the example spectrograms in Figure \ref{fig:generated_images}. In the final example for walking, ACGAN generated a spectrogram which has a gap over which the micro-Doppler is nearly zero. Kinematically, this corresponds to a situation in which the subject was walking, took two steps (as evidenced by the two peaks in frequency), stopped, and then took another two steps walking forward.     

\subsection{Kinematic Fidelity}
The patterns observed in radar spectrograms directly relate to kinematics of the motion being observed. For example, in the case of walking,  the torso response represents the strongest return and exhibits a sinusoidal oscillation.  The periodic motion of the legs causes the highest frequency oscillations around the main Doppler shift.  Known as physical features, such properties have often been used in classification of micro-Doppler signatures.  In this section, we consider kinematic properties of synthetic walking and falling signatures, as these are challenge cases for the ACGAN due to the great diversity within real training samples as well as the greater richness of frequencies comprising the signature.

	
	

In particular, kinematic fidelity of the synthetic signatures are evaluated by imposing upon the images three different kinematic rules:
\begin{enumerate}
    \item Generated spectrograms should be periodic, and thus represent the cyclic motion of the body.
    \item The maximum torso frequency should be lower than the that of the legs.
    \item If the generated spectrogram occupies the positive frequencies, indicating that the motion is performed towards the radar, the signature should not contain any high negative frequency components, and vice versa.
\end{enumerate}
 The first rule is only applicable to periodic motions, whereas the other two rules can be applied to walking and falling. Note that there is no guarantee that these kinematic rules ensure that every generated signature is fully compatible with the kinematic constraints of human motion.  However, they do serve to enforce the most basic properties of the skeletal constraints on human motion, and can eliminate unrealistic or impossible synthetic signatures.  
 
\begin{figure}[t!]
\centering
\includegraphics [width=3.5in] {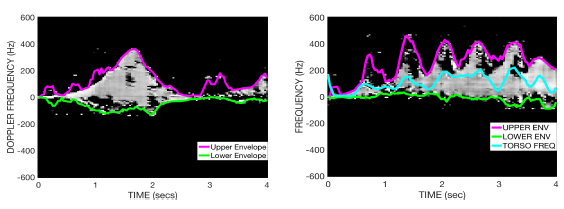}
\caption{Rule definition process: low-pass filtered upper/lower envelope and torso frequency extraction in the generated images (left: falling, right: walking).}
    \label{fig:rules}
\end{figure}

\begin{figure}[t!]
\centering
\includegraphics [width=3.5in] {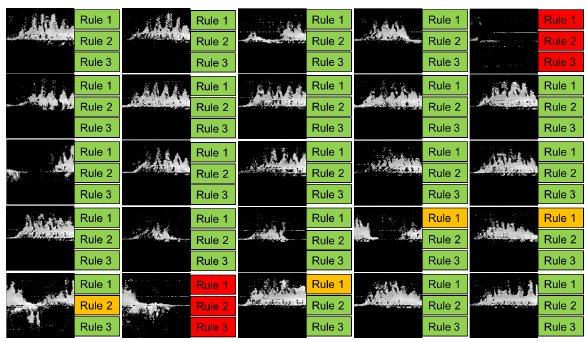}
\caption{Output of the kinematic sifting algorithm on the 25 randomly selected walking spectrograms generated by ACGAN. Green, orange, and red colors indicate that image passes the rule without any problems, passes the rule with minor problems, and fails the rule.}
	\label{fig:Kinematic_Fig_2}
\end{figure}

 The above three rules can be tested by extracting the upper/lower envelopes and the torso frequency, as depicted in Figure \ref{fig:rules}. To illustrate the process of sifting the data with these kinematic rules, let us randomly select 25 synthetic walking spectrograms generated by ACGAN, as shown in Figure \ref{fig:Kinematic_Fig_2}.   The green labels indicate that the synthetic images passed all three kinematic rules, while orange indicates minor issues (i.e. only one or two rules failed), and red indicates that the image fails.  Inspecting the two images from this random selection of 25 signatures, it may be observed that one fails because it only has a faint clutter component and essentially no target component.  The other signature that failed all rules overtly has no periodicity and inconsistent distribution of positive and negative frequencies. Furthermore, the strongest response is not consistent with typical torso motion.
 
 When these rules are applied to the 5,000 synthetic walking spectrograms generated by ACGAN, 15\% of the signatures failed all three of the kinematic rules.  For falling, only the 2nd and 3rd rules were enforced, resulting in the failure of 10\% of the signatures.  These results mean that while ACGANs are predominantly successful in simulating human motion, there is nevertheless a significant portion of the synthetic data which is kinematically impossible and could lead to a degradation in classification accuracy.  In this work, we propose implementation of a PCA-based kinematic sifting algorithm to eliminate such undesirable synthetic samples prior to utilizing the synthetic data for training.  This is discussed in more detail in Section V.
 
\subsection{Synthetic Data Diversity}
A generative model is considered unsuccessful if it only outputs one type of image (also known as mode collapse). This is a well-known phenomenon in GANs, where the generator will collapse and outputs a single prototype that maximally fools the discriminator \cite{wgan}. To evaluate potential mode collapse, we utilized a quantitative similarity measure called MS-SSIM \cite{wang_image_2004}. MS-SSIM attempts to discount aspects of an image that are not important for human perception. It assumes the values range between [0, 1] where higher values correspond to perceptually more similar images, and smaller values indicate a better diversity. Mathematically, it is defined 
\begin{equation}
    \mathrm{SSIM}(\mathrm{x},\mathrm{y}) = [l_M(\mathrm{x},\mathrm{y})]^{\alpha_M} \prod_{j=1}^M [c_j(\mathrm{x}, \mathrm{y})]^{\beta_j}[s_j(\mathrm{x}, \mathrm{y})]^{\gamma_j},
\end{equation}
where $\alpha_M$, $\beta_j$ and $\gamma_j$ are used to adjust relative importance of different components. Luminance, contrast and structure comarison measures are defined as $l((\mathrm{x},\mathrm{y}))$, $c((\mathrm{x},\mathrm{y}))$, $s((\mathrm{x},\mathrm{y}))$, respectively. $M$ depicts the Scale number that will be used in the iterative filtering and downsampling. $\mathrm{x}$ and $\mathrm{y}$ are defined as the compared images.

\begin{figure}[t!]
\centering
\includegraphics [width=3.5in] {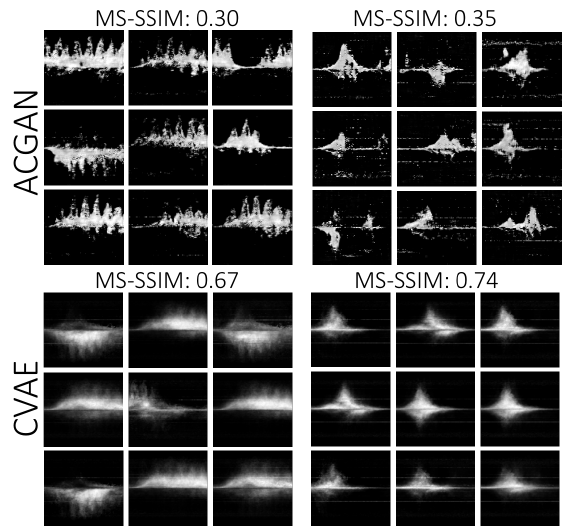}
\caption{MS-SSIM scores for randomly chosen 100 walking (left) and falling (right) image pairs for ACGAN (top) and CVAE (bottom).}
\label{fig:Diversity_Fig_1}
\end{figure}

As a simple example, we randomly selected 100 image pairs from both CVAE and ACGAN-based synthetic spectrogram datasets of the walking and falling classes.  Some sample images from the chosen pairs can be seen in Figure \ref{fig:Diversity_Fig_1}.  The MS-SSIM values for CVAE for walking and falling are found to be 0.67 and 0.74, respectively. These values indicate that CVAE has low diversity for the selected random pairs. For the ACGAN, MS-SSIM values are found to be 0.30 for walking, and 0.35 for falling, indicating a much higher degree of diversity among the synthetic signatures generated.  For comparison, measured samples of falling and walking yielded MS-SSIM values of 0.45 and 0.40, respectively. Thus, the ACGAN generated signatures provide not only sharper images, relative to CVAE generated signatures, but also a greater degree of diversity that is comparable to that expected based on measured data.  

\begin{figure*}[t!]
\centering
\includegraphics [width=7in] {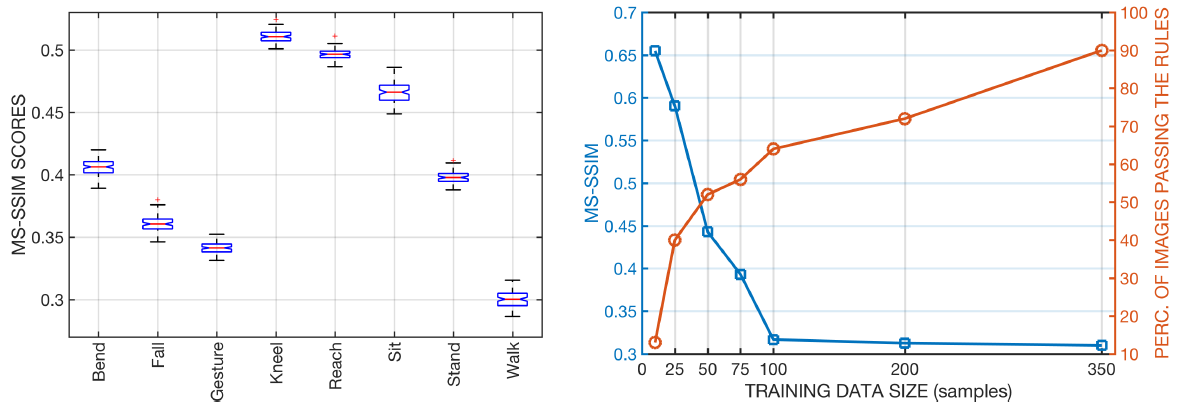}
\caption{Diversity measures: (a) Box plots of the intra-class diversities measured by MS-SSIM, (b) MS-SSIM diversity values and percentages of kinematically correct images as a function of the real training samples used in the training.}
\label{fig:Diversity_Fig_2}
\end{figure*}

The results of detailed analysis of the MS-SSIM values for ACGAN-generated synthetic data are given in Figure \ref{fig:Diversity_Fig_2}, which shows a box plot of the MS-SSIM values for 100 randomly selected image pairs in each class.  The walking class exhibits the most diversity, as would be expected by the possible variations of a complex motion.  Gesturing and falling exhibit the next greatest levels of diversity.  Considering that falling is a more or less uncontrolled motion, and that gesture is highly open to participant interpretation during enactment, these results match expectations.

The level of diversity in the synthetic dataset generated also depends upon the amount of measured data provided to ACGAN during generation. Figure \ref{fig:Diversity_Fig_2}-(b) depicts the variation of MS-SSIM values for falling and walking as a function of training data size.  Juxtaposed on top of this curve is the relation between training data size and percentage of synthetic samples that pass the kinematic rules.  Note that when a minuscule amount of measured data is used to drive the ACGAN, the network has a tendency to generate a large amount of data that is highly similar (over 0.8 MS-SSIM values), and that are also kinematically meaningless - virtually none of the synthetic signatures actually adheres to kinematic rules.  It is only when at least 100 measured samples are supplied the ACGAN that signatures are obtained which predominantly meet kinematic rules and exhibit a high degree of diversity.  Further increasing the amount of measured data supplied for training the ACGAN does not significantly affect the data diversity, but does increase the kinematic fidelity of the data generated.  Note that when 350 measured training samples are utilized, the percentage of data passing the kinematic rules rises to 90\% - a 25\% increase of the percentage attained with just 100 measured training samples.  

In micro-Doppler literature, studies involving several thousand measurements are typical.  The proposed method, however, requires only 350 samples to generate 40,000 synthetic samples.  This represents a 10 fold decrease in data collection requirements, while enabling increased sample diversity and a 100-fold increase in the size of the training dataset.  As shown in Section V, DNNs trained on this synthetic dataset outperforms DNNs trained on measured data only.

\begin{figure}[t]
\centering
\includegraphics [width=3.5in] {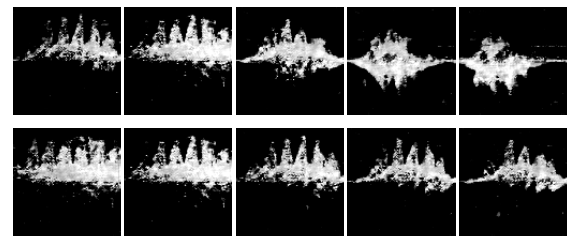}
\caption{Strolling on the latent space: top row  and bottom rows depict the generated images by changing the first and second latent variables to (-3.0, -1.5, 0, 1.5, 3.0), respectively. }
\label{fig:latent_1}	
\end{figure}

\subsection{Strolling in the Latent Space}

	

Analysis of the latent manifold helps us to understand the model details, indicates the signs of memorization, and shows if the latent space is hierarchically collapsed \cite{radford_unsupervised_2015}. Kinematic and physical changes (such as different velocities, the orientation of the target, the direction of the motion, etc.) in the generated spectrograms while strolling through the latent space indicate that the model has learned relevant and interesting representations from the training data. As an example, consider an ACGAN retrained with a generator that has a latent size of $5\times1$. After the training is complete, a walking image is generated by randomly sampling the latent variables from a uniform distribution and passing it through the generator. Then, we changed the first latent variable value between -3.0 and 3.0 with linear increments of 1.5. Note that the other four latent variables are kept fixed during this operation. The resulting walking spectrograms for different latent variable values are depicted in the top row of Figure \ref{fig:latent_1}. 

Examining the top row of Figure \ref{fig:latent_1}, as the value of the latent variable is increased, the Doppler bandwidth is first reduced and then begins to flip, with an increasing peak in the negative Doppler frequencies.  Moreover, the Doppler bandwidth does indeed vary according to the aspect angle between the radar line-of-sight and target direction of motion.  Thus, it may be deduced that the first latent variable models direction of motion.

Next, we change the second latent variable and observe the resulting changes in micro-Doppler (see second row of Figure \ref{fig:latent_1}.  It is evident that this variable models stride rate.  This may be seen by counting the peaks in the signature.  While the leftmost spectrogram has six distinct peaks, with each peak corresponding to a step, the last spectrogram on the right has only three peaks.  This indicates that stride rate decreases as the second latent variable increases.

%

\begin{figure}[t!]
\centering
\includegraphics [width=3.3in] {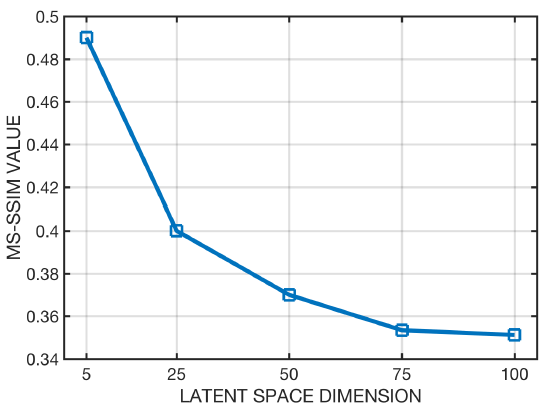}
\caption{The ACGAN latent space analysis of falling spectrograms in terms of diversity measurements.}
\label{fig:Latent_falling}	
\end{figure}

As can be seen from the above example, the dimensionality of the latent space effectively relates to how the network models the underlying representation of the data. In general, the question of how many latent variables should be used in GANs still remains unanswered. However, it is known that the real distribution arises out of lower dimensional latent distributions. There is also a concern if a lower dimensional latent space is utilized, the GAN might not have enough information to model the data, causing the modes to collapse. A large latent space dimension, on the other hand, makes the model so complex that the training time becomes overly long. Moreover, the mapping of latent variables into spectrograms becomes difficult in high dimensional latent space.  Some earlier works have used 100 as a de facto value \cite{radford_unsupervised_2015}.  Using this as a baseline, we examine the effect of latent space dimensionality on the MS-SSIM values of resulting synthetic spectrograms.  Five ACGAN models are trained with different latent space dimensions: 5, 25, 50, 75 and 100. The resulting MS-SSIM diversity metrics for each model is shown in Figure \ref{fig:Latent_falling} for the falling class. It may be seen that small latent space dimensions suffer from low diversity due to the limited number of combinations that can be achieved, while increasing the latent space dimension yields better diversity. However, beyond 75 latent variables, the diversity starts to plateau, indicating that increasing complexity is offering little benefit to data diversity. Thus, in this work we elect to use 100 latent variables in our implementation of the ACGAN.

%



\subsection{Evaluation of Kinematic Fidelity with PCA}
In this section, we propose a kinematic sifting algorithm using generalized PCA (GPCA) \cite{erol_realization_2018}. Kinematic evaluation of the synthetic spectrograms in Section IV proved that some spectrograms are still kinematically not consistent with true human motion. The proposed sifting algorithm aims to eliminate some of the inconsistent images that might degrade classification performance. GPCA is first applied on the real training images for each class, $D_i, i=1, 2, ..., 8$. The objective is to find a matrix subspace set ${\mathrm{\tilde{U}}_{D_i}^{(1)} \in \mathbb{R}^{I_1 \times P_1}}$ and ${\mathrm{\tilde{U}}_{D_i}^{(2)} \in \mathbb{R}^{I_2 \times P_2}}$ that project the original tensor into a low dimensional matrix subspace $\mathrm{Y}_m^{D_i} \in \mathbb{R}^{P_1 \times P_2}$ (with $P_1 \leq I_1$ and $P_2 \leq I_2$) defined as

\begin{equation}
\mathrm{Y}_m^{D_i} = \mathrm{S}_m^{D_i} \times_1 \mathrm{U}_{D_i}^{(1)^T} \times_2 \mathrm{U}_{D_i}^{(2)^T},
\end{equation}
where $\mathrm{S}_m^{D_i}$ is the real training spectrogram from class $D_i$. The objective function of the GPCA can be written as
\begin{equation}
(\mathrm{\tilde{U}}_{D_i}^{(1)},\mathrm{\tilde{U}}_{D_i}^{(2)}) = \underset{\mathrm{U}^{(1)},\mathrm{U}^{(2)}}{\text{arg max}} \sum_{m=1}^M \big\| \mathrm{Y}_m^{D_i} - \overline{\mathrm{Y}^D_i} \big\|^{2}_{F},
\end{equation}
where $\mathrm{\overline{Y}} = \frac{1}{M} \sum_{m=1}^M \mathrm{Y}_m$. The core matrix for each $m$ samples can be obtained by projecting the original images using optimized subspaces, $\mathrm{\tilde{U}_{D_i}^{(1)},\tilde{U}_{D_i}^{(2)}}$, as

\begin{equation}
\mathrm{\tilde{Y}}_m^{D_i} = \mathrm{S}_m^{D_i} \times_1 \mathrm{\tilde{U}}_{D_i}^{(1)^T} \times_2 \mathrm{\tilde{U}}_{D_i}^{(2)^T}.
\end{equation}

Finally, the feature vector of a training sample for a specific class, $m$, can be constructed as $\mathrm{C}_{m} = \mathrm{vec}\big(\mathrm{\tilde{Y}_m}\big), \ \in \mathbb{R}^{1 \times D}$, where $D= P_1\times P_2$ and $\mathrm{vec}(\cdot)$ is the matrix column-wise vectorization operator. We defined $P_1$ and $P_2$ as 2. 

Upon finding the optimized subspaces and reduced feature space for each class, the $n$-dimensional convex hull method is applied to determine the feature space boundaries of the each class. Falling and walking feature space boundaries (with feature space dimensions set to 3) are shown in Figure 10. Next, the optimized subspaces on the synthetically generated images are used to reduce the dimensionality of the feature space. Finally, features are checked to ensure they fall within the specific class boundaries, as determined by real training examples. Pseudcode of the proposed method is depicted in algorithm 2. By using the sifting method we are able to eliminate 11\% of bending, 18\% of falling, 8\% of gesture, 33\% of kneeling, 7\% of reaching, 15\% of sitting, 36\% of standing, 22\% of walking. This elimination reduces the generated dataset size from 40,000 to 31,133.

\section{ACGAN with PCA-based Kinematic Sifting}
\begin{figure}[t!]
\centering
\includegraphics [width=3.6in] {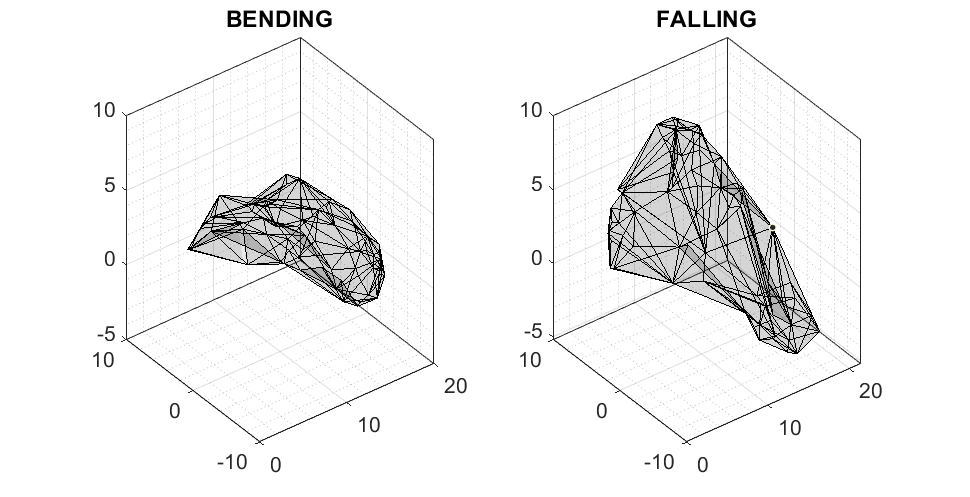}
\caption{Boundaries determined by n-dimensional convex hull for bending and falling.}
    \label{fig:Testing_prp_1}
\end{figure}
\begin{algorithm}[b!]
	\caption{Data sifting with generalized PCA}
	\begin{algorithmic}[1]
		\renewcommand{\algorithmicrequire}{\textbf{Input:}}
		\renewcommand{\algorithmicensure}{\textbf{Output:}}
		\REQUIRE Real and generated spectrograms  \\
		\ENSURE  Kinematically sifted images
		\FOR {EACH CLASS}
		\STATE  Read real spectrograms
		\STATE  Apply GPCA subspace learning method and find the optimized subspaces, (100x100 spectrogram dimensions reduced to 2x2)
		\STATE  Find the boundaries of the reduced feature space using the 4 dimensional convex hull method
		\STATE  Save the convex hull parameters and optimized subspaces.
		\ENDFOR
		\FOR {EACH CLASS}
		\STATE Read the genered spectograms
		\STATE Load the optimized subspaces and convex hull parameters
		\STATE Apply optimized subspaces and get the reduced feature space
		\STATE Check if the current generated spectrogram is within the convex hull boundaries with a tolerance
		\STATE If in keep else eliminate
		\ENDFOR
		\RETURN Sifted images
	\end{algorithmic}
\end{algorithm}

\begin{table*}[t!]
	\caption{Performance comparison (accuracy) between TF-AlexNet, CVAE, and ACGAN, PCA-ACGAN with tolerance 1.0 and PCA-ACGAN with tolerance 0.5.}
	\label{tab:acctable}
	\centering
	\begin{tabular}{*{8}{l}}

		\toprule
		&  TF-AlexNet & TF-VGG16 & CVAE  & ACGAN & PCA-ACGAN-TOL-1.0 & PCA-ACGAN-TOL-0.5 \\
		\midrule
		Accuracy 	   & 0.765 & 0.842 & 0.732 & 0.825 & 0.877   & 0.932  \\

		\bottomrule 

	\end{tabular}
	
\end{table*}

\begin{table*}[t!]
	\centering
	\caption{Test confusion matrix for PCA-ACGAN-DCNN with tolerance 0.5 (test accuracy of 93\%).  }
	\label{tab:confmat7class}
	\begin{tabular}{|
			>{\columncolor[HTML]{C0C0C0}}c |c|c|c|c|c|c|c|c|c|}
		\hline
		\textbf{\%}      & \cellcolor[HTML]{C0C0C0}\textbf{Bending}                    & \cellcolor[HTML]{C0C0C0}\textbf{Falling}                    & \cellcolor[HTML]{C0C0C0}\textbf{Gesture}                    & \cellcolor[HTML]{C0C0C0}\textbf{Kneeling}                   & \cellcolor[HTML]{C0C0C0}\textbf{Reaching}                   & \cellcolor[HTML]{C0C0C0}\textbf{Sitting}                    & \cellcolor[HTML]{C0C0C0}\textbf{Standing}                   &
		\cellcolor[HTML]{C0C0C0}\textbf{Walking}                        \\ \hline
		\hline
		\textbf{Bending} & \cellcolor[HTML]{32CB00}{\color[HTML]{333333} \textbf{100}} & 0& 0 & 0  &0   &0
		 &
		0                                                            & 0                                                             \\ \hline
		
		\textbf{Falling} & 0 & \cellcolor[HTML]{32CB00}{\color[HTML]{333333} \textbf{90}}                                                           & 3                                                            & 0                                                           & 2                                                           &
		0
		&
		                                  0                          & 5                                                             \\ \hline
		
		\textbf{Gesture} &  0 & 0                                                           & \cellcolor[HTML]{32CB00}{\color[HTML]{333333} \textbf{96}}                                                            & 0                                                           & 0                                                           &
		0 &
		0                                                            & 4                                                             \\ \hline
		
		\textbf{Kneeling} & 20 & 0                                                           & 0                                                            & \cellcolor[HTML]{32CB00}{\color[HTML]{333333} \textbf{80}}                                                           & 0                                                           &
		0 &
		0                                                            & 0                                                             \\ \hline
		
		\textbf{Reaching} & 0 & 0                                                           & 0                                                            & 0                                                           & \cellcolor[HTML]{32CB00}{\color[HTML]{333333} \textbf{84}}                                                          &
	16 &
		0                                                            & 0                                                             \\ \hline
		
		\textbf{Sitting} &0 & 0                                                          & 0                                                            & 2                                                           & 0                                                          &
		\cellcolor[HTML]{32CB00}{\color[HTML]{333333} \textbf{98}}  &
		0                                               & 0                                                             \\ \hline
		
		\textbf{Standing} & 0 & 0                                                           & 0                                                            & 2                                                           & 0                                                           &
		0 &
		\cellcolor[HTML]{32CB00}{\color[HTML]{333333} \textbf{98}}                                                            & 0                                                             \\ \hline
		\textbf{Walking} & 0 & 0                                                           & 0                                                            & 0                                                           & 0                                                           &
		0 &
		0                                                            & \cellcolor[HTML]{32CB00}{\color[HTML]{333333} \textbf{100}}                                                             \\ \hline
	\end{tabular}
\end{table*}

\subsection{Experimental Results}
\subsubsection{Classification Accuracy}
In this section, we present classification performances of transfer learning on AlexNet (TF-ALexNet) and VGGnet (TF-VGG16), DCNNs trained on the synthetic spectrograms generated by CVAE (CVAE-DCNN) and ACGAN (ACGAN-DCNN) with and without kinematic sifting at various tolerances. Note that both AlexNet and VGG16 are pre-trained on ImageNet. After the weights of the networks are acquired, only the last 2 layers are retrained using the real radar data. Moreover, the last softmax layer was also adjusted to the number of classes.  Moreover, we tested two different tolerances in the proposed kinematic sifting algorithm, labeled as PCA-ACGAN-TOL-1.0 and PCA-ACGAN-TOL-0.5. 

To analyze the improvement achieved by the generative models, we collected a challenging data test set in a completely different environment and configuration from the real samples (mentioned in Section II), which were used to train CVAE and ACGAN. The test performances are presented in \ref{tab:acctable} in terms of accuracy. The average test accuracies for TF-AlexNet, TF-VGG16, CVAE-DCNN, ACGAN-DCNN, PCA-ACGAN-DCNN-TOL-1.0 and PCA-ACGAN-DCNN-TOL-0.5 are determined to be 76\%, 84\%, 73\%, 82\%, 87\%, and 93\%, respectively. In prior studies, VGGnet is a network that has provided accuracies that have surpassed that of other pre-trained networks, such as GoogleNet, as well as convolutional autoencoders (CAEs) and supervised learning with handcrafted features \cite{SeyfiogluCAE_TAES}, when the amount of training data is limited \cite{seyfioglu_deep_2017}.  Therefore, it is not suprising that VGGnet surpasses the performance of AlexNet, and even that attained by using the unsifted, initial training database generated by ACGAN without consideration of any kinematics.  The low performance of the CVAE-DCNN is also expected since the generated images are blurry and unrealistic, have low diversity, and there was no fine tuning in the training.  

Significantly, the performance of the DCNN \textit{increases} as the ACGAN-generated synthetic signatures are increasingly sifted, identifying and discarding those samples that are kinematically impossible and therefore un-representative of the related class label.  The most sifting is done with the smallest tolerance, and the performance of the proposed PCA-ACGAN-DCNN-TOL-0.5 approach is drastically higher than that achieved with transfer learning, or training data that is unsifted.  This result further demonstrates the need to consider physics and kinematics in the generation of synthetic training data for micro-Doppler classification.

\subsubsection{Confusion Matrix}

The test confusion matrix of the PCA-ACGAN-DCNN-TOL-0.5 is provided in \ref{tab:confmat7class}. It is observed that the proposed scheme provides the best test accuracy around 93\%. The primary source of confusion is between bending and kneeling, as expected. These two motions have the same kinematic structure in the TF domain. Both include a positive and a negative hump adjacent to each other, which depict the motion of the upper body in forward direction for reaching, and the motion of the upper body in the downwards direction for kneeling. One significant difference between these activities, however, is the motion of the knee.  In some experiments the motion of the knee was very pronounced, resulting in increased confusion.

There is also a high misclassification rate between reaching and sitting. Some reaching signatures only contain the forward or backward motion of the upper body, which results in similar signatures as that for sitting.  Next, there is 5\% misclassification between falling and walking. This is caused by the presence of progressive falls in the training database and lower stride rates in some of the testing signatures. Some testing samples include only one or two strides meaning that subject only took 1 or 2 steps. These signatures have similar visual representation to falling signatures due to their low periodicity. 

A high fall detection performance (90\%) is achieved by employing the proposed method.  Note that, experimental results are based on real TWR radar data obtained from multiple aspect angles (0$^{\circ}$, 30$^{\circ}$, 45$^{\circ}$, 60$^{\circ}$, 90$^{\circ}$), which demonstrates that the proposed algorithm yields the highest overall classification accuracy among other methods.

\begin{figure}[t!]
\centering
\includegraphics [width=3.5in] {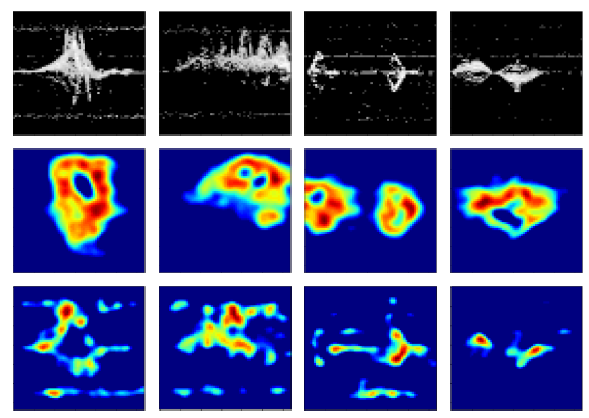}
\caption{Original TWR (first row) spectrograms and corresponding saliency maps achieved by the ACGAN-DCNN (second row) and DCNN (third row) for different class samples (columns left to right: falling, walking, gesture, reaching).}
\label{fig:sal}	
\end{figure}

\begin{figure*}[t!]
\centering
\includegraphics [width=5.5in] {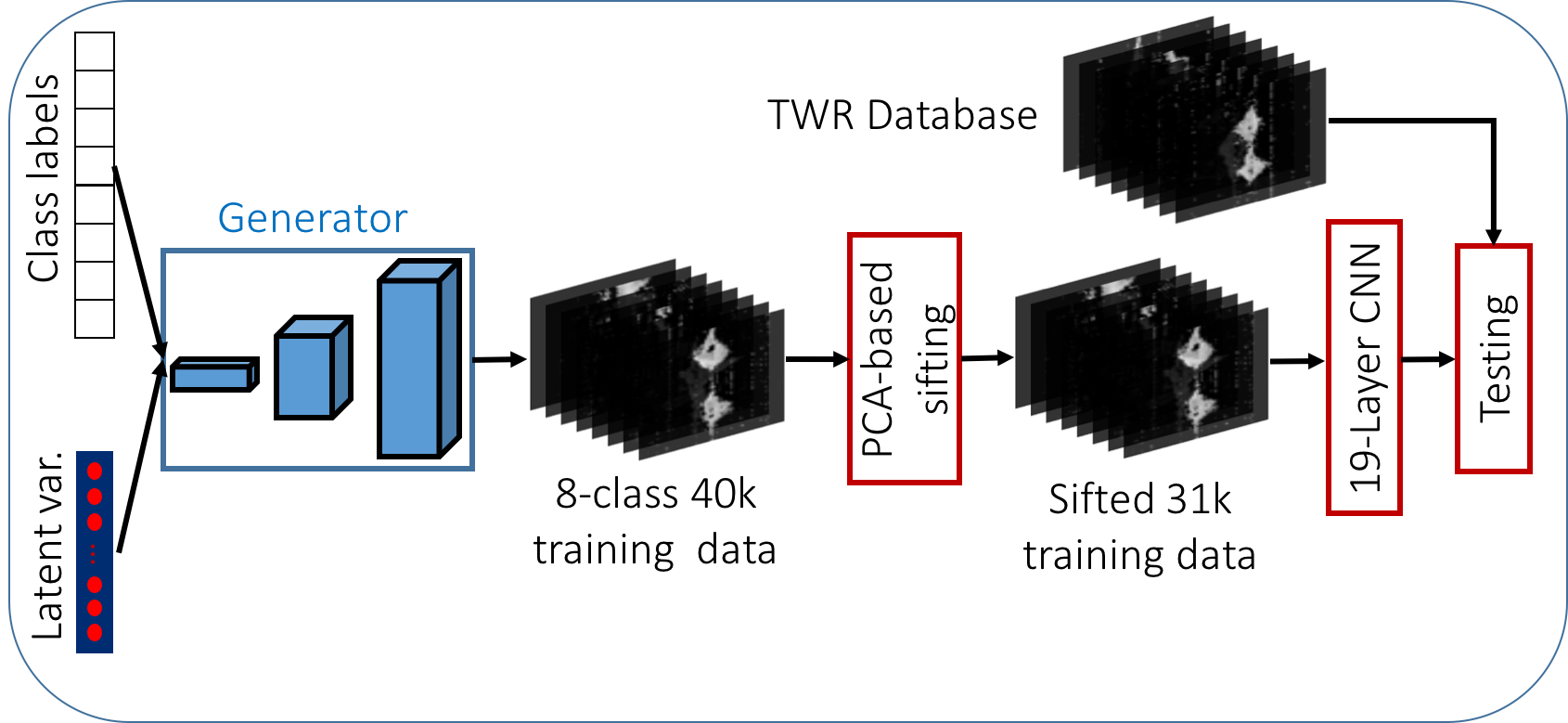}
\caption{Flow chart of the proposed approach with PCA sifting algorithm. Note that, dataset 2 was collected in a TWR setup.}
    \label{fig:Testing_prp_1}
\end{figure*}

\subsubsection{Saliency maps}
The benefits of training the DCNN with a large synthetic training dataset can also be illustrated by examining saliency maps of the network. A saliency map is an image that shows each pixel's unique importance according to the DCNN's classification declaration \cite{simonyan_deep_2013}. Saliency maps of four test spectrogram images are computed for ACGAN-DCNN and DCNN (trained with dataset 1) and are shown in Figure \ref{fig:sal}. It may be observed that the ACGAN-DCNN places more importance on the perimeter of the actual signatures in the spectrograms, and ignores the noisy parts in the images.  In contrast, a  basic CNN trained on measured data focuses on some of the noisy parts and looks for specific pixels rather than the signature envelope. More specfically, the ACGAN-DCNN wraps all physical components of the falling, whereas the basic CNN puts importance on the highest frequency components. This trivialization by the basic CNN can be problematic with high aspect angle data. For the "reaching" class, the ACGAN saliency map shows that the network tries to connect two different (positive and negative) components together. whereas those components are disjoint in the basic CNN.  In summary, the ACGAN-DCNN approach enables correct identification of the motion components of the spectrogram, rejecting clutter components.  Saliency map observations thus reinforce the importance of training on a large dataset that has great diversity, while also maintaining kinematic fidelity.

\section{Conclusion}

In this paper, we proposed a novel approach for generating synthetic radar micro-Doppler signatures for human motion classification.  The proposed approach leverages auxilliary conditional generative adversarial networks (ACGANs) to build a diverse dataset for training deep neural networks. However, the ACGAN-generated signatures include kinematically impossible signatures, which can degrade classification performance.  To overcome this problem, a PCA-based kinematic sifting algorithm was proposed to eliminate inconsistent samples that could corrupt DNN training.  A 19-layer DCNN trained on kinematically sifted ACGAN-based synthetic data was shown to be effective in classifying challenging datasets collected across different environments, as demonstrated by 93\% correct classification. In our experiment, test data were collected through-the-wall, while LOS measurements were used to drive the ACGAN in training data generation.  This result surpasses other previously proposed approaches, including transfer learning and ACGAN-generated data that is not kinematically sifted.

\ifCLASSOPTIONcaptionsoff
  \newpage
\fi

\bibliographystyle{IEEEtran}
\bibliography{GenModels}



%

%
\begin{IEEEbiography}[{\includegraphics[width=1in,height=1.25in,clip,keepaspectratio]{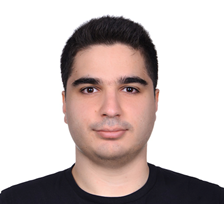}}]{Baris Erol}

received the B.S. degree in electrical and electronics engineering and the MSc. degree in electrical engineering from the TOBB University of Economics and Technology, Ankara, Turkey, in 2014 and 2015 respectively, and the Ph.D. degree in electrical and computer engineering from Villanova University, Villanova, PA, USA, in 2018. He currently works as a research scientist at Siemens Corporate Technology in the Automation Runtime Systems research group, Princeton, NJ, USA. His expertise is in sensor integration, radar signal processing, and machine learning in which he has published more than 20 peer-reviewed articles. 
\end{IEEEbiography}

\begin{IEEEbiography}[{\includegraphics[width=1in,height=1.25in,clip,keepaspectratio]{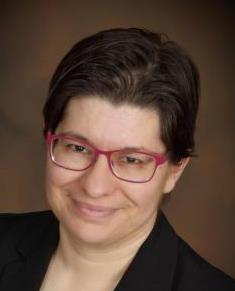}}]{Sevgi Z. Gurbuz}
(S'01-M'10-SM'17) received the B.S. degree in electrical engineering with minor in mechanical engineering and the M.Eng. degree in electrical engineering and computer science from the Massachusetts Institute of Technology, Cambridge, MA, USA, in 1998 and 2000, respectively, and the Ph.D. degree in electrical and computer engineering from Georgia Institute of Technology, Atlanta, GA, USA, in 2009.  
    
From February 2000 to January 2004, she worked as a Radar Signal Processing Research Engineer with the U.S. Air Force Research Laboratory, Sensors Directorate, Rome, NY, USA.  Formerly an Assistant Professor in the Department of Electrical-Electronics Engineering at TOBB University, Ankara, Turkey and Senior Research Scientist with the TUBITAK Space Technologies Research Institute, Ankara, Turkey, she is currently an Assistant Professor in the University of Alabama at Tuscaloosa, Department of Electrical and Computer Engineering.  Her current research interests include radar signal processing, machine learning and pattern recognition, cognitive radar, and sensor networks.
     
Dr. Gurbuz is a recipient of the 2020 SPIE Rising Researcher Award, EU Marie Curie Research Fellowship, USAF Achievement Medal, USAF Commendation Medal, AFRL Technical Achievement Award, National Defense Science and Engineering Fellowship, C.S. Draper Fellowship, and the 2010 IEEE Radar Conference Best Student Paper Award.

\end{IEEEbiography}

\begin{IEEEbiography}[{\includegraphics[width=1in,height=1.25in,clip,keepaspectratio]{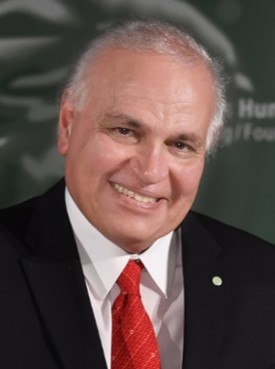}}]{Moeness G. Amin}
(F'01) received the Ph.D. degree in electrical engineering from the University of Colorado, Boulder, CO, USA, in 1984.  Since 1985, he has been with the Faculty of the Department of Electrical and Computer Engineering, Villanova University, Villanova, PA, USA, where he became the Director of the Center for Advanced Communications, College of Engineering, in 2002. 

Dr. Amin is a Fellow of the Institute of Electrical and Electronics Engineers; Fellow of the International Society of Optical Engineering; Fellow of the Institute of Engineering and Technology; and Fellow of the European Association for Signal Processing. Dr. Amin is the Recipient of the 2017 Fulbright Distinguished Chair in Advanced Science and Technology; Recipient of the 2016 Alexander von Humboldt Research Award; Recipient of the 2014 IEEE Signal Processing Society Technical Achievement Award; Recipient of the 2009 Individual Technical Achievement Award from the European Association for Signal Processing; Recipient of the 2015 IEEE Aerospace and Electronic Systems Society Warren D. White Award for Excellence in Radar Engineering; Recipient of the IEEE Third Millennium Medal; Recipient of the 2010 NATO Scientific Achievement Award; Recipient of the 2010 Chief of Naval Research Challenge Awrd; Recipient of the Villanova University Outstanding Faculty Research Award, 1997; and the Recipient of the IEEE Philadelphia Section Award, 1997. He was a Distinguished Lecturer of the IEEE Signal Processing Society, 2003-2004, and was a member then the Chair of
the Electrical Cluster of the Franklin Institute Committee on Science and the Arts, 2000-2015. Dr. Amin has over 800 journal and conference publications in signal processing theory and applications. He co-authored 22 book chapters and is the Editor of the three books Through the Wall Radar Imaging, Compressive Sensing for Urban Radar, and Radar for Indoor Monitoring published by CRC Press in 2011, 2014, and 2017, respectively.
\end{IEEEbiography}







\end{document}